\documentstyle[12pt]{article}

\begin{document}

\author{F. Despa  \\ 
Department of Theoretical Physics\\Institute of Atomic Physics \\
PO Box MG - 6, Magurele - Bucharest, Romania\\e-mail: despa@theor1.ifa.ro}
\title{Jellium Correction on the Critical Condition of Cluster Fission 
within a Liquid Drop Model}
\date{Z. Phys. {\bf D} - to appear}
\maketitle

\begin{abstract}
Jellium correction on the fissionability
parameter is estimated within a Liquid Drop Model of the charged metallic
cluster. This correction modifies the critical condition of fission and, 
it becomes relevant for small  multicharged clusters.  \\ 
\centerline {{\bf PACS numbers:} 36.40.Qv, 36.40.-c}
\end{abstract}
Adapting the Liquid Drop Model $(LDM)$ from nuclear physics to the
description of finite systems (metallic clusters) \cite{1,2} or even
infinite ones (metals) \cite{3,4} is straight-forward. The physical insight
of the model is explained in the following way \cite{5} : In a system with
positive surface tension and positive curvature energy, the energy minimizes
when the system has an environment that is as bulklike as possible, i.e.,
when the surface is as small in area and as concave as it can be for a given
volume. Thus, the total energy of an extended system of volume $V$ and
surface area $S$ can be readily expressed as a sum of volume, surface,
electrostatic (for charged systems) and curvature terms.

Despite the errors arising from the shell structure of finite systems, $LDM$
can be successfully used in understanding the fission of multiply charged
metallic clusters \cite{6}.
Neglecting the curvature energy (with
supposition on a small curvature of the droplet), the total energy of the
charged droplet is assumed to be comprised of three terms in this approach, 
\begin{equation}
E_T=E_b+E_S+E_C.
\end{equation}
The leading term of (1) may be written $E_b=\varepsilon \cdot N$, where $N$
is the number of constituent particles and $\varepsilon $ is the energy per
particle in the limit $\frac VS\rightarrow \infty $. 
When a spherical droplet of radius $R_o$ is deformed (conserving
the volume) toward an ellipsoidal shape its area changes to 
\begin{equation}
S \approx 4\pi R_o^2\left( 1+\frac{2}5\beta ^2\right) ,
\end{equation}
upon the second-order approximation in powers of the deformation parameter 
$\beta $. 

The surface energy, $E_s$, 
in this the simplest deformation of the droplet is therefore 
\begin{equation}
E_{S} \approx 4\pi \sigma R_o^2\left( 1+\frac{2}5\beta ^2\right) ,
\end{equation}
where $\sigma $ is the surface tension. 
Also, in the simplest manner, when the deformed droplet electrified
to a charge $q$, it has an electrostatic (Coulomb) energy 
\begin{equation}
E_{C} \approx \alpha \frac{q^2}{4 \pi \epsilon R_o} \left( 1 - \frac{1}{5}
\beta ^2 \right). 
\end{equation}
Here, $\alpha = \frac{3}{5} $ for a uniform
distribution of charge $q$ inside the droplet or $\alpha = \frac{1}{2} $
when the charge $q$ is spread over the surface only. Looking at eqs. (3 - 4)
one can observe that the relative strengths of the Coulomb and surface
forces will determine the stability of the spherical shape and hence its
tendency to fission.

The Coulomb energy of a charged metallic cluster is a long - standing
controversial problem in cluster physics. Thus, the dimensionless cvoefficient
$\alpha$ (see eq. 4) takes experimental values around 0.4 \cite{7} while
from the classical electrostatic reasons, two different theoretical results
for this coefficient have been derived: $\alpha = \frac{1}{2}$ from the 
spherical-
capacitor approach \cite{8}, and $\alpha = \frac{3}{8}$ from the 
image-potential approach \cite{9}. 
Recently, Seidl and Perdew \cite{10} have established that,
classically, $\alpha = \frac{1}{2}$ by either approach and the experimental
deviation from this value must be accounted for by quantum effects. Their
theoretical result is derived in a "point-charge-plus-continuum" model for
jellium within which the excess electric charge is uniformly distributed
in a surface shell between the radii 
${\left(R^3_o - r^3_s \right)}^{\frac{1}{3}}$ and $R_o$.
In this model, $\alpha = \frac{1}{2}$ is a correct result in the limit
$z \cdot \frac{r_s^3}{R_o^3} \ll 1$ which means large clusters and not so
many electric charges ($z \cdot e$).
Actually, according with the above model, $\alpha $ is larger
than $\frac{1}{2}$ depending on the charge $z \cdot e$ and the number of the 
constituent particles. This follows from the fact that, increasing the cluster
charge or decreasing the number of the cluster constituents,
the volume of the charged surface shell increases relative to that of 
all the cluster volume and, consequently,
$\alpha \rightarrow \frac{3}{5}$ which is the right value for uniformly 
charged sphere. This assumption can be relevant for small  multicharged
clusters. With these in mind somewhat corrections on the critical condition
of cluster fission within a {\it Liquid Drop Model (LDM)\/}
are necessary. This is, briefly, the aim for that the present paper is 
addressed.

In the "Wigner - crystal" model of jellium, the positive charge of the ions
is replaced by a rigid uniform positive background with the charge density $
+e \overline{n} = e \frac{3}{4 \pi r_s^3}$ inside the sphere of radius $R_o$.
The valence electrons are point charges sitting at equilibrium position
inside the sphere of radius $R_o$. Sometimes, by various reasons,
the cluster ionization, the above model is replaced by the
"point - charge - plus - continuum" model of jellium. In this approach, the
electrons to be removed remain point charges while, the other electrons are
replaced by a continuous, mobile fluid of negative charge. This system
achieves equilibrium when the net electric field vanishes wherever the
density of continuum negative charge does not. In this way, when the point
electrons are still located inside the droplet, the density of continuum
negative charge must neutralize the positive background everywhere but,
inside the spheres of radius $r_s$ around the point electrons the density of
continuum negative charge must vanish. As the electrons are pulled out
through the surface of the cluster their "classical holes" deform, remaining
behind inside the surface while conserving its volume, after the electrons
have been removed to $x = \infty$, these "classical holes" are spread over a
shell of charge density $+e \overline{n}$ between the radii $R_o$ and 
$R = {\left( R_o^3 - z r_s^3 \right)}^{\frac{1}{3}}$, where $z$ is the number 
of the electrons which have been removed. In this way, the Coulomb potential,
subject of the Laplace equation both inside the neutral part of the cluster 
$(0 < r < R)$ and outside the cluster $(r > R_o)$ and of the Poisson equation
in the cluster region between $R_o$ and $R$, reads 
\begin{equation}
\Phi (r) = \cases {A, &if $r<R$;\cr
- \frac{e \overline{n}}{6 \epsilon}r^2 - \frac{B}{r} + C, &if $R<r<R_o$;
\cr
- \frac{D}{r}, &if $r>R_o.$\cr} 
\end{equation}
Imposing appropriate boundary conditions, the A,B,C and D constants are
readily determined and, the electrostatic potential in the charged shell 
$(R<r<R_o)$ is 
\begin{equation}
{\Phi}_{shell} = - \frac{e \overline{n}}{6 \epsilon}r^2 - \frac{R^3}{3
\epsilon r} e \overline{n} + \frac{e \overline{n}}{6 \epsilon R_o} \left(
R_o^3 + 2R^3 \right) - \frac{e \overline{n}}{3 {\epsilon}_o R_o} \left(R^3 -
R_o^3 \right), 
\end{equation}
where ${\epsilon}_o$ is the electric permittivity of free space.
Under the assumption that $\epsilon = {\epsilon}_o$, the electrostatic 
self-energy reads 
\begin{eqnarray}
W &=& \frac{1}{2} \int dV e \overline{n} \cdot {\Phi}_{shell} \\ \nonumber
&=& \frac{1}{2} \cdot \frac{(e)^2}{4 \pi \epsilon R_o} \cdot 
\frac{3}{5 \frac{r_s^6}{R_o^6}} \cdot
\left( 2 - 5 \left( 1- z \frac{r_s^3}{R_o^3} \right) + 
3 {\left( 1 - z \frac{r_s^3}{R_o^3} \right)}^{\frac{5}{3}} \right).
\end{eqnarray}
As we can see, for $ z = 1 $, the above amount is the electrostatic 
self-energy obtained by Seidl and Perdew ( equation 3 in \cite{10}).
Further, we will note $\frac{r^3_s}{R_o^3}$ by $\eta$ and,
in the second-order approximation in powers of this
parameter, the above equation becomes 
\begin{equation}
W = \frac{{(ze)}^2}{ 4 \pi {\epsilon} R_o} \cdot \frac{1}{2} \left( 1 + 
\frac{\eta}{9} + \frac{{\eta}^2}{27} + \cdots \right),
\end{equation}
which means that the value of the $\alpha$ parameter having
corrections from the jellium approach of the metallic cluster reads 
\begin{equation}
\alpha = \frac{1}{2} \left( 1+ \frac{\eta}{9} + \frac{{\eta}^2}{27} + \cdots
\right) 
\end{equation}
and, it is easy to see that when $\eta \rightarrow 0$ (the charged shell
tends to an infinitesimal thin layer) then $\alpha \rightarrow \frac{1}{2}$
and, when $\eta \rightarrow 1$ (the charged surface shell has an increasing
trend by various reasons: small clusters and/or multicharged clusters)
then $\alpha \rightarrow \frac{3}{5}$. 
On the other hand, the radius of the spherical cluster can be expressed
as $R_o = r_s N^{\frac{1}{3}}$ which means that $\eta = \frac{z}{N}$ and 
$\alpha$ becomes 
\begin{equation}
\alpha = \frac{1}{2} \left( 1+ \frac{z}{9N} + \frac{z^2}{27N^2} + \cdots
\right),
\end{equation}
which shows that the dimensionless parameter $\alpha$ is larger than
$\frac{1}{2}$ depending on the charge z and the number of cluster
constituents N. 

As we have said, the above correction on the value of the dimensionless
parameter $\alpha$ may have somewhat relevance on the critical condition
of cluster fission within the $LDM$.

For a droplet $N^{z+}$, where $N$ is the number of atoms and $z$ is the
charge in units $e$, the surface energy is proportional to $N^{\frac{2}{3}}$
and is a minimum when the droplet is spherical. The surface energy disfavors
deformation (see eq. 3). On the other
hand, the Coulomb energy depends on $N^{- \frac{1}{3}}$ and is a maximum
when the droplet is undeformed (see eq. 4); it favors the deformation. 
The relative strengths of the Coulomb and surface forces will
determine the stability of the shape and hence its tendency to fission.
Looking at equations (3 - 4) and, taking into account the last equation for
the $\alpha$, one can say that the droplet is unstable to (classical)
spontaneous decay, even at zero temperature, when 
\begin{eqnarray}
& &\frac{z^2}{N} \left( 1+ \frac{z}{9N} + \frac{z^2}{27N^2} + \cdots \right)
\rightarrow
{\left(\frac{z^2}{N} \left( 1+ \frac{z}{9N} + \frac{z^2}{27N^2} + 
\cdots \right)
\right)}_c = \\ \nonumber
& &= \frac{64 {\pi}^2 \epsilon r_s^3 \sigma}{e^2}.
\end{eqnarray}
Looking at the above equations we may say that the corresponding 
critical size, $N_c$, for $\alpha$
given by (10) slightly differs from that corresponding to 
$\alpha = \frac{1}{2}$.
The jellium correction rises the lower bound of the critical sizes.

With high enough excitation, fission is always an allowed decay
channel. Whether fission is observed or not, however, depends on where the
threshold energy for competing reactions lies in relation to the fission
threshold. Evaporation is the main competitor in both nuclear and cluster
fission. 
Looking, again, at the above equations one may see that it is expected a 
slight decreasing of the fission
barrier for $\alpha$ given by (10) relatively to that for 
$\alpha = \frac{1}{2}$.

As a concluding remark, one may say that the above jellium correction on the
fissionability criterion within a $LDM$, apart from the fact that promotes a
correct (natural) way to account for the Coulomb energy as a part in the
fission condition, this rises (indeed not very much) the lower bound of
the critical sizes as a consequence of $\alpha $ increasing and, on the other
hand, decreases the fission barrier by the same considerations.\newline
The author thanks to Professor M. Apostol for many useful discussions.


\newpage

\begin{thebibliography}{99}
\bibitem{1}  Sugano S., A. Tamura and Y. Ishii: Z. Phys. {\bf D12}, 213
(1989).

\bibitem{2}  Lipparini E. and A. Vitturi: Z. Phys. {\bf D18}, 222 (1990).

\bibitem{3}  Perdew J. P., Y. Wang and E. Engel: Phys. Rev. Lett {\bf 66},
508 (1991).

\bibitem{4}  Perdew J. P., P. Ziesche and C. Fiolhais: Phys. Rev. {\bf B47},
16460 (1993).

\bibitem{5}  Rowlinson J.S. and B. Widom: {\it Molecular Theory of
Capillarity \/}, Clarendon, Oxford, (1982).

\bibitem{6}  Saunders W.A. : Phys. Rev. Lett. {\bf 64}, 3046 (1990); {\bf 66},
840 (1991); W. A. Saunders and N. Dam: Z. Phys. {\bf D20}, 111 (1991).

\bibitem{7}  de Heer W.A. : Rev. Mod. Phys. {\bf 65}, 611, (1993);
Kappes M.M.: Chem. Rev. {\bf 88}, 369, (1988).

\bibitem{8}  
Herrmann A., E. Schumacher and L. Woste: J. Chem. Phys. {\bf 68}, 2327, (1978).

\bibitem{9}  
Smith J.M.: Am. Inst. Aeronaut. Astronaut. J. {\bf 3}, 648, (1965); Wood D.M.:
Phys. Rev. Lett. {\bf 46}, 749 (1981).

\bibitem{10} Seidl M. and J.P. Perdew: Phys. Rev. {\bf B50},
5744, (1994).

\end{thebibliography}
\end{document}